\documentclass[amsmath,amssymb, aps, pra,reprint,footinbib,floatfix,superscriptaddress]{revtex4-2}

\usepackage{float}
\usepackage{amsmath,amssymb}
\usepackage[english]{babel}
\usepackage{upgreek}
\usepackage{dsfont}
\usepackage{graphicx}
\usepackage{color}
\usepackage{xcolor}
\usepackage{bbold}
\usepackage[hidelinks]{hyperref}
\usepackage[authormarkup=none]{changes}
\definechangesauthor[name={Stefan}, color=purple]{s}
\definechangesauthor[name={Bjoern}, color=blue]{b}
\definechangesauthor[name={Klaus}, color=red]{k}

\newcommand{\V}{{\cal V}}
\newcommand{\diff}{{\rm d}}

\begin{document}

\title{Testing collapse models with Bose-Einstein-Condensate interferometry}

\author{Bj\"{o}rn Schrinski}
\affiliation{University of Duisburg-Essen, Faculty of Physics, Lotharstra\ss e 1, 47048 Duisburg, Germany}
\affiliation{Center for Hybrid Quantum Networks (Hy-Q), Niels Bohr Institute, University of Copenhagen, Blegdamsvej 17,
DK-2100 Copenhagen, Denmark}
\author{Philipp Haslinger}
\affiliation{Vienna Center for Quantum Science and Technology, Atominstitut, TU Wien, Stadionallee 2, 1020 Vienna, Austria}
\author{Jörg Schmiedmayer}
\affiliation{Vienna Center for Quantum Science and Technology, Atominstitut, TU Wien, Stadionallee 2, 1020 Vienna, Austria}
\author{Klaus Hornberger}
\affiliation{University of Duisburg-Essen, Faculty of Physics, Lotharstra\ss e 1, 47048 Duisburg, Germany}
\author{Stefan Nimmrichter}
\affiliation{Naturwissenschaftlich-Technische Fakult\"at, Universit\"at Siegen, Walter-Flex-Stra\ss{}e 3, 57068 Siegen, Germany}

\begin{abstract}
The model of continuous spontaneous localization (CSL) is the most prominent consistent modification of quantum mechanics predicting an objective quantum-to-classical transition. Here we show that precision interferometry with Bose-Einstein condensed atoms can serve to lower the current empirical bound on the localization rate parameter by several orders of magnitude. This works by focusing on the atom count distributions rather than just mean population imbalances in the interferometric signal of squeezed BECs, without the need for highly entangled GHZ-like states. In fact, the interplay between CSL-induced diffusion and dispersive atom-atom interactions results in an amplified sensitivity of the condensate to CSL.
We discuss experimentally realistic measurement schemes utilizing state-of-the-art experimental techniques to test new regions of parameter space and, pushed to the limit, to probe and potentially rule out large relevant parameter regimes of CSL.

\end{abstract}

\maketitle

\section{Introduction}\label{Introduction}

Postulating an objective, spontaneous collapse process for the wave function is a way to overcome the quantum measurement problem and to explain the fundamental absence of spatial superpositions on the macroscopic scale \cite{bassi2013models}. This idea deems quantum mechanics incomplete and complements it with a fundamental stochastic modification that bridges the gap between the  micro-cosmos of quantum phenomena and the classical world. 

A prime example is the model of continuous spontaneous localization (CSL) \cite{bassi2013models,ghirardi1990markov}, which predicts a mass-amplified spatial decoherence effect. It reinstates macrorealism \cite{leggett2002testing} and can be motivated from natural consistency requirements on generic  `classicalizing' modifications of quantum mechanics \cite{Nimmrichter2013}.
The spontaneous collapse is  accompanied by a tiny amount of diffusive heating,  impacting also classical states of motion, which could however be mitigated by adding colored noise \cite{adler2007collapse,adler2008collapse} and friction \cite{smirne2015dissipative} to the model. 

\begin{figure}
  \centering
  \includegraphics[width=0.45\textwidth]{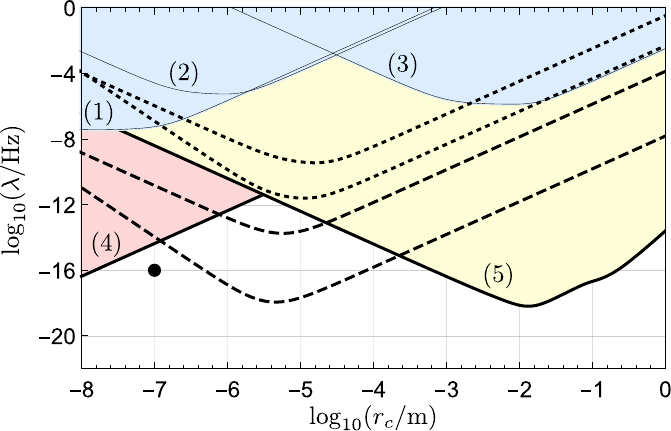}
  \caption{CSL bounds from different experiments. The blue-shaded region represents the falsified parameters from the three most macroscopic interferometric tests so far: (1) near-field interferometry \cite{fein2019quantum}, (2) MZI with atoms \cite{xu2019probing}, and (3) MZI with BEC \cite{asenbaum2017phase}. The red-shaded region marks the best non-interferometric bounds from (4) monitoring spontaneous x-ray emission \cite{Arnquist2022Search}, while the yellow shaded region marks bounds from classical force-noise sensing in (5) the LISA pathfinder mission \cite{armano2016sub,carlesso2016experimental}. 
The here proposed MZI setups in double and single wells are associated with dotted and dashed exclusion curves, respectively, determined by the parameters in Tab.~\ref{tab:parameters}. The dot marks historically chosen values for collapse parameters  \cite{ghirardi1986unified,ghirardi1990markov}.}
\label{fig:CSL_exclusion}
\end{figure}

The CSL hypothesis has sparked numerous efforts to conceive \cite{Nimmrichter2011,bateman2014near,Laloe2014,Bahrami2014,
Nimmrichter2014,Diosi2015,schrinski2017collapse,bilardello2017collapse,stickler2018probing,bahrami2018testing,pino2018chip,
tilloy2019neutron,forstner2020nanomechanical} and perform \cite{Kovachy2015,carlesso2016experimental,piscicchia2017csl,fein2019quantum,vinante2020challenging,pontin2020ultranarrow,vinante2020testing} experiments that rule out a significant portion of its two-dimensional parameter space comprised of the CSL localization rate $\lambda$ and the localization length scale $r_C$. Each experimental test falsifies a certain set of parameters marked by an exclusion curve $\lambda(r_C)$. The best experimental bounds so far \cite{Arnquist2022Search,carlesso2016experimental} are surveyed in Fig.~\ref{fig:CSL_exclusion} (solid lines). They do not yet reach the critical regime of nano- to micrometer localization lengths and CSL rates as low as the historic value $\lambda=10^{-16}\,$Hz for the reference mass $1\,$u at $r_C=100\,$nm (black dot) \cite{ghirardi1986unified,ghirardi1990markov}. 

Accessing this regime with a \textit{quantum} experiment could refute the CSL hypothesis, especially since non-classical CSL tests are robust against the aforementioned model extensions \cite{adler2007collapse,adler2008collapse,smirne2015dissipative}. 
However, the most macroscopic matter-wave experiments to date (thin solid lines) are still many orders of magnitude away. A purpose-built space mission would have to demonstrate interference of a $10^9\,$u microparticle over more than $100$ seconds \cite{kaltenbaek2012macroscopic,kaltenbaek2016macroscopic}, a  challenging endeavour.
The state-of-the-art CSL bounds are obtained from classical noise measurements on optomechanical sensors probing the CSL-induced heating (thick solid lines)
\cite{carlesso2016experimental,carlesso2018non,vinante2020challenging}, which are most sensitive to CSL at larger length scales, and from searches for anomalous x-ray emission \cite{Arnquist2022Search}, which are most sensitive at shorter $r_C$.

We instead consider collective spin states in a standard Mach-Zehnder interferometer (MZI) setting where a Bose-Einstein-Condensate (BEC) with  $N$ atoms of mass $m$ is coherently  split into two spatially separate arms and recombined after an effective interrogation time $t$ at a second beam splitter.
Experimental realizations include double-well trapping of condensates on a chip \cite{Berrada2013}, atoms suspended in optical standing-wave antinodes \cite{xu2019probing}, and free-falling momentum-split condensates \cite{asenbaum2017phase}.

The article is structured as follows: In Sec.~\ref{Standard two-mode BEC interferometry} we recapitulate the standard Mach-Zehnder interference scheme in the collective spin representation and discuss both the usual measurement analysis in terms of interferometric visibility and the direct analysis of the atom count distribution. In Sec.~\ref{Mathematical model}, we present our quantitative model for the time evolution of an interfering two-mode BEC state in the presence of CSL, and we proceed to solve it under the assumption of many particles by performing a continuum approximation in Sec.~\ref{sec:dispDiff}. This is followed by the key result of this paper: we propose to estimate the range of excluded CSL parameters directly from the measured atom-count distribution and investigate how the accuracy of the estimate scales with the particle number in the BEC in Sec.~\ref{ScalingProp}. Remarkably, we find that Mach-Zehnder schemes with spatially overlapping arms offer superior scaling and could probe the most relevant CSL parameter regime with available technology. We substantiate our finding with 
a proper statistical analysis based on the Fisher information in Sec.~\ref{Statistics}, taking into account unavoidable noise sources like two- and three-body recombinations. Finally,  we briefly discuss the impact of further modified CSL models on the discussed setups in Sec.~\ref{Generalizations of the CSL model} and give some conclusive remarks in Sec.\ref{Conclusion}.

\section{Standard two-mode BEC interferometry}\label{Standard two-mode BEC interferometry}

Consider a Mach-Zehnder interference scheme in which a BEC of $N$ atoms is uniformly and coherently split into two modes by a beam splitter, freely evolved in time, and then recombined at another beam splitter, after which one records the atom counts in the two output ports. 
The standard measurement protocol varies the mean interferometric phase $\bar\varphi$ characterizing the effective path difference between the interferometer arms and extracts the interference visibility $\V$ from the mean count difference $\bar{n} = N \V \sin \bar{\varphi}$ upon recombination. 
Given that the measured visibility is always lower than the ideal value $\V = 1$ predicted by quantum theory, 
bounds on the CSL parameters could be obtained by attributing the uncontrolled visibility loss to spontaneous collapse. However, previous in-depth studies showed that a thus defined CSL test offers no collective advantage over single-atom interferometry \cite{bilardello2017collapse,schrinski2020quantum}, regardless of any initial squeezing \cite{schrinski2019macroscopicity}: 
for non-interacting BEC and spatially separate arms, CSL predicts $\V_{\rm CSL} = \exp (-\Gamma_{\rm P} t/2)$, with a \emph{single-atom} dephasing rate $\Gamma_{\rm P}/2 = (m/{\rm u})^2 \lambda f_\mathrm{P}(r_C)$. 
Equating the CSL value with the observed $\V$ divides the CSL parameter space into an excluded ($\V_{\rm CSL} \leq \V$) and a compatible region, subject to statistical error analysis.
The geometry factor $f_\mathrm{P}$ assumes its maximum $f_\mathrm{P}(r_C)=1$ when $w_x \ll r_C \ll \Delta_x$, given the separation $\Delta_x$ between the interferometer arms and the spatial extension $w_x$ of the interfering modes \cite{schrinski2019macroscopicity}. For much greater or smaller $r_C$, it scales like $(\Delta_x/r_C)^2$ and $(r_C/w_x)^2$, respectively, if we assume an effectively one-dimensional condensate elongated along the $z$-direction (with $w_y \approx w_x \ll w_z$). A 
tightly confined atom cloud ($w_x \approx w_y \approx w_z$)
would yield $f_\mathrm{P}(r_C) \sim (r_C/w_x)^3$ for $r_C \ll w_x$.

\begin{figure*}
  \centering
  \includegraphics[width=0.89\textwidth]{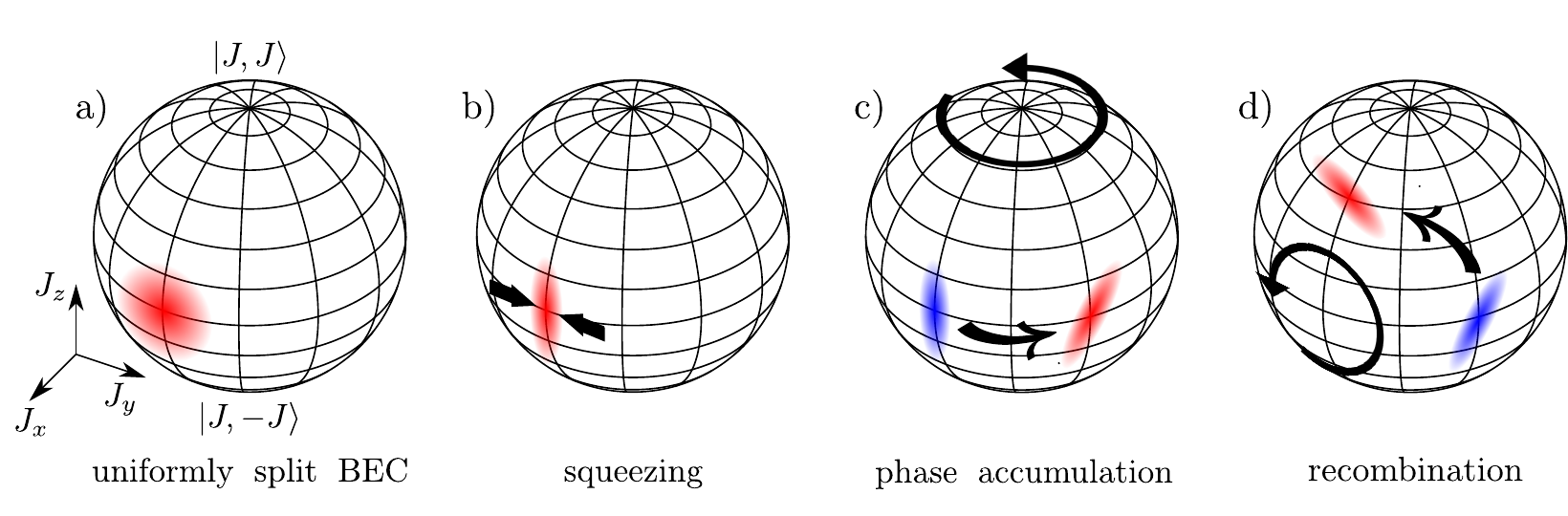}
  \caption{Sketch of a two-mode BEC interference scheme in the collective spin picture, as represented by the Husimi-$Q$ function on the generalized Bloch sphere \cite{Ma2011}. (a) A uniformly split BEC in a coherent spin state 
  corresponds to a symmetric Gaussian peak localized on the equator, as depicted by red shading on the left. 
  The azimuthal angle corresponds to the relative phase between the two modes, while the polar angle characterizes the
  population imbalance. (b) Typically, the initial superposition state is first squeezed azimuthally
 to increase phase sensitivity. (c) The subsequent accumulation of a relative phase by propagation through the two interferometer modes corresponds to a rotation around the vertical axis by the respective  angle. 
 (d) The phase distribution may then be detected by another $\pi/2$-rotation around 
 an equatorial axis, as set by the recombining beam splitter. This converts the phase spread to a spread in population differences, as seen in the atom count statistics.
  }
\label{fig:RamseyScheme}
\end{figure*}

In this article, instead of focusing on 
derived quantities like interference visibility, we will study the impact of CSL on the phase distribution of the condensate. Consider a uniformly split BEC, as characterized by an initial particle number distribution with vanishing mean population imbalance and a number uncertainty $\sigma_n (0)$. The conjugate many-body quantum observable $\varphi$ has an uncertainty  bounded by $\sigma_\varphi (0)>1/\sigma_{n}(0)$ \cite{Ma2011}. The mean value $\bar \varphi$ accrues over time depending on the energy difference between the two modes.
In each run, the value of $\varphi$ could be read out directly by overlapping and imaging the phase-sheared condensates corresponding to the two modes \cite{sugarbaker2013enhanced}. More conventionally, it is mapped to the atom number differences $n$ between the modes by recombination at the second beam splitter ($\pi/2$-pulse) \cite{Berrada2013}, see Fig.~\ref{fig:RamseyScheme} for a sketch of a typical protocol in the collective spin representation \cite{Arecchi1972}. 

Any increase of the initial phase spread $\sigma_\varphi$ caused by dephasing channels such as CSL would yield a decreased interference visibility $\V$ after recording many runs of the protocol with varying phase $\bar \varphi$. Such a variation, however, is not necessary. One can
directly sample the $\varphi$-distribution, and extract the width $\sigma_\varphi(t)$, by measuring the atom counts and recording a histogram over many experimental runs at one chosen $\bar \varphi$-value, without the need to record an entire interference pattern.

For $N$ indistinguishable bosons in two modes, represented by annihilation operators $\mathsf{c}_{a,b}$, we can make use of the collective spin representation \cite{Arecchi1972}, with $J=N/2$ and spin operators $\mathsf{J}_x=(\mathsf{c}_a^\dagger\mathsf{c}_b+\mathsf{c}_b^\dagger\mathsf{c}_a)/2$, $\mathsf{J}_y=(\mathsf{c}_a^\dagger\mathsf{c}_b-\mathsf{c}_b^\dagger\mathsf{c}_a)/2i$, and $\mathsf{J}_z=(\mathsf{c}_a^\dagger\mathsf{c}_a-\mathsf{c}_b^\dagger\mathsf{c}_b)/2$.
Dual Fock states $|n_a , N-n_a \rangle$ in the two modes correspond to the $\mathsf{J}_z$-eigenstates $|J,M\rangle$ with $M = n_a-N/2$ denoting half the atom number difference. An unsqueezed, uniformly split BEC with a mean relative phase $\bar \varphi$ corresponds to a product state of $N$ single-atom superpositions, $|\psi \rangle \propto (\mathsf{c}^\dagger_a+e^{i\bar\varphi}\mathsf{c}^\dagger_b)^N|\mathrm{vac}\rangle$, also known as a coherent spin state. Formally, it is obtained from the spin state $|J,J\rangle$ by performing a $\pi/2$ rotation around a spin axis perpendicular to $\mathsf{J}_z$, as can be represented pictorially in terms of the Husimi-Q function on the generalized Bloch sphere \cite{Ma2011}. We make use of this function to visualize the interferometric schemes in Figs.~\ref{fig:RamseyScheme} and \ref{fig:EchoSchemes}. 

In the standard protocol of Fig.~\ref{fig:RamseyScheme}, the initially uncorrelated superposition state (at $\bar \varphi = 0$) is first phase-squeezed  before it accumulates a mean interferometric phase $\bar \varphi$. The squeezing can be realized, for example, by means of a one-axis twisting operation \cite{Kitagawa1993},
\begin{align}\label{eq:OneaxisSqueezing}
\mathsf{U}_\xi = \exp (- i\eta_\xi \mathsf{J}_x) \exp (i\chi_\xi \mathsf{J}_z^2). 
\end{align}
It describes a shearing operation of strength $\chi_\xi$ along the equator, followed by a rotation by $\eta_\xi$ around $\mathsf{J}_x$, chosen such that the resulting state exhibits the variance $\xi^2/N$ along the equator at the desired value of the squeezing parameter $\xi$. After the squeezing step, the state accumulates a mean phase as it rotates around the $\mathsf{J}_z$-axis of the Bloch sphere. In the presence of atom-atom interactions, this would be accompanied by an additional shearing rate $\zeta$, which we will consider in the next section. 

Finally, upon readout, the condensate state is rotated by $\pi/2$ around a fixed spin axis perpendicular to $\mathsf{J}_z$, which maps the state's phase distribution to the distribution of population differences, and vice versa. The probability to measure a certain number difference $n$ in the output port is given by $p(n|\lambda; r_C, I)=|\langle J,n/2|\rho_t|J,n/2\rangle|^2$, which can be approximated for large $N$ with Jacobi theta functions \cite{schrinski2019macroscopicity}. Here, the background information $I$ subsumes other experimental parameters that influence the result. For values $\bar{\varphi}$ not too close to $\pm\pi/2$ and a well-defined phase right before recombination, the measured count distribution simplifies further to a Gaussian, whose variance is determined by the state's phase variance before recombination, \eqref{eq:phaseVar_dephasing} or \eqref{eq:phaseVar_withDiff}, multiplied by $N^2\cos^2\bar{\varphi}$. This transformation is not necessary if the phase is read out directly \cite{sugarbaker2013enhanced,schrinski2020quantum}.

\begin{figure}
  \centering
  \includegraphics[width=0.45\textwidth]{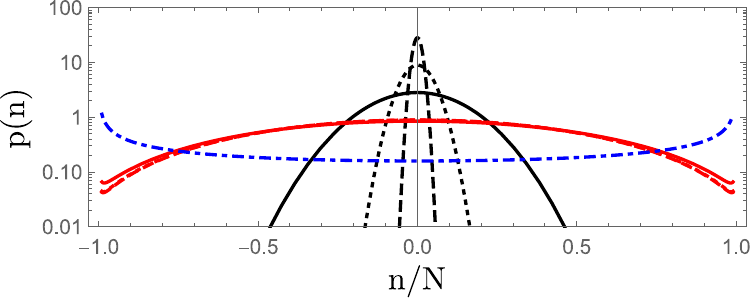}
  \caption{Impact of phase noise on the atom count distribution in a standard two-mode BEC interferometer with an initially unsqueezed split condensate of various atom numbers $N$. We plot the distribution of the atom number difference $p(n)$ between the two output ports at $\bar{\varphi}=0$  against the normalized difference $n/N$, assuming condensates with $N=10^2$ (solid), $10^3$ (dotted), and $10^4$ (dashed) atoms. The black curves represent the ideal case without any loss or decoherence, whereas the corresponding red curves (which all coincide) result from an added phase uncertainty $\sigma^2_{\rm{conv}} = 0.1$ before recombination. For comparison, the dashed-dotted blue curve represents a completely dephased state.
   }
\label{fig:Gaussians}
\end{figure}

Before we introduce the dynamical model of the CSL effect, let us remark on the benefit of  condensates over single atoms. Why is it better to sample the number distribution of an $N$-atom condensate in $k$ measurement runs than to perform $kN$ runs of single-atom interferometry? The answer, detailed in Secs.~\ref{ScalingProp} and \ref{Statistics}, lies in the initial phase uncertainty $\sigma_\varphi^2 (0) \propto 1/N$, 
which can be squeezed below shot noise. Thus, in the regime of 
number-resolving and phase-stable precision interferometry, the atom-count distribution of the condensate becomes highly sensitive to collective phase-broadening disturbances such as CSL.
These do not act on each individual atom independently, but on the entire condensate, and they are further amplified in interplay with atom-atom interactions.
Unfortunately, the collective scaling advantage is often lost under practical circumstances, whenever conventional noise sources smear out the phase significantly, $\sigma_\varphi^2 \to \sigma_\varphi^2 + \sigma_{\rm{conv}}^2$ (thereby diminishing interference visibility notably below $1$). This is illustrated in Fig.~\ref{fig:Gaussians}, which compares the ideal (decoherence-free) shot noise limited count distributions after recombination for various $N$-values (black curves) to the respective cases with added $\sigma^2_{\rm{conv}} = 0.1$ (red). (The corresponding interference visibility would be $\V \approx 0.9$.) The distributions are dominated by conventional  smearing and thus no longer distinguishable, making the case for precision experiments with as little noise as possible to boost CSL sensitivity.

We remark that the illustrated dephasing behavior, whether caused by conventional noise or CSL, differs \emph{fundamentally} from the respective outcome with $N$ individual atoms. In the limit of strong dephasing, the latter scenario would result in a 50:50 chance for each atom to end up in one of the detectors: the exact opposite behavior of a condensate with a macroscopic, strongly fluctuating phase \cite{schrinski2020quantum}. Only in the ideal case without decoherence do the statistics of condensate and single-atom interferometry agree; and squeezing increases the discrepancy further.

\section{Two-mode description of CSL}\label{Mathematical model}

We will now show how the time evolution of the two-mode condensate in the presence of CSL can be modelled theoretically. We include atom-atom interactions and allow for spatially overlapping modes, as will be crucial for achieving enhanced  sensitivity on the collapse parameters. 

In the presence of CSL, the time evolution of the collective spin state representing the condensate can be described by the master equation \cite{schrinski2019macroscopicity}
\begin{align}\label{eq:LioSup}
\partial_t\rho=\frac{1}{i\hbar}[\epsilon\mathsf{J}_z+\hbar\zeta\mathsf{J}_z^2,\rho]+\mathcal{L}\rho .
\end{align}
The first term in the Hamiltonian describes the free rotation around the $z$-axis of the generalized Bloch sphere at an angular frequency given by the energy difference $\epsilon$ between the two involved modes. The second term accounts for atom-atom interactions in the condensate to lowest order close to the equator, $\hbar \zeta=2 (\diff \mu/ \diff n )_{n=0}$, where 
$\mu$ is the chemical potential \cite{Javanainen1997PhaseDispersion}. 

Spontaneous collapse 
contributes the Lindblad generator \cite{schrinski2019macroscopicity}
\begin{align}\label{eq:TwoModeLindblad}
\mathcal{L}\rho=&\lambda\frac{ m^2}{\rm u^2}\,\frac{r_{C}^3 }{\pi^{3/2}} \int \diff^3 {q}\, e^{-q^2r_C^2}\nonumber\\
&\times\left(\mathsf{A}(\mathbf{q})\rho\mathsf{A}^\dagger(\mathbf{q})-\frac{1}{2}\left\{\mathsf{A}^\dagger(\mathbf{q})\mathsf{A}(\mathbf{q}),\rho\right\}\right),
\end{align}
with $\mathsf{A}(\mathbf{q})=\int d^3{p}\,\mathsf{a}^\dagger(\mathbf{p})\mathsf{a}(\mathbf{p-\hbar q})$ and $\mathsf{a} (\mathbf{p})$ the particle annihilation operator in momentum representation (with dimensions of a wave number). 
Restricting the dynamics to the two relevant modes of the interferometer, we can expand the Lindblad operator as
\begin{align} \label{eq:TwoModeLindbladOp}
\mathsf{A}(\mathbf{q}) \simeq& \sum_{j,k \in \{a,b\}} \underbrace{\langle \psi_j|e^{i \mathbf{q}\cdot\boldsymbol{\mathsf{r}} }|\psi_k \rangle}_{=:W_{jk} (\mathbf{q})} \mathsf{c}_j^\dagger \mathsf{c}_k \\
=&\left[ W_{aa}(\mathbf{q}) + W_{bb}(\mathbf{q}) \right] \frac{N}{2} + \left[ W_{aa}(\mathbf{q}) - W_{bb}(\mathbf{q}) \right] \mathsf{J}_z \nonumber \\
& + \left[W_{ab} (\mathbf{q}) + W_{ba} (\mathbf{q}) \right] \mathsf{J}_x + i \left[W_{ab} (\mathbf{q}) - W_{ba} (\mathbf{q}) \right] \mathsf{J}_y. \nonumber 
\end{align}
Here, $|\psi_{a,b}\rangle$ denote the single-particle wave functions associated with the two modes, $\boldsymbol{\mathsf{r}}$ the respective position operator, and $W_{jk} (\mathbf{q}) = W_{kj}^{*} (-\mathbf{q})$ the matrix elements of the momentum displacement operator. The $\mathsf{J}_z$-term will cause incoherent phase flips and thus be responsible for CSL-induced dephasing, while the $\mathsf{J}_x$-term causes atoms to hop between the two modes and thereby induces diffusion. The $\mathsf{J}_y$-term vanishes for bound states with real-valued wave functions and will therefore be ignored.

The terms omitted in \eqref{eq:TwoModeLindbladOp} would describe incoherent hopping of atoms between the condensate and other undetected modes, which causes atom loss. Formally, we can account for this CSL-induced depletion effect in the data analysis of an experiment by introducing conditional outcome probabilities, given that the detected $N$ atoms have remained in the condensate at all times. This way, the CSL bounds would not depend on the additional classical observation of depletion \cite{schrinski2019macroscopicity}. Moreover, since particles lost from the condensate (or, less likely, regained ones) will always increase the phase uncertainty $\sigma_\varphi$, our omission of the depletion effect only underestimates the influence of CSL. 

Plugging the Lindblad operator \eqref{eq:TwoModeLindbladOp} into the master equation \eqref{eq:TwoModeLindblad} yields the combined effect of CSL-induced dephasing (random phase flips) and diffusion (spin flips), 
\begin{align}
\mathcal{L}\rho &\simeq \Gamma_\mathrm{P}\left(\mathsf{J}_z\rho\mathsf{J}_z-\frac{1}{2}\left\{\mathsf{J}_z^2,\rho\right\}\right)+
\Gamma_\mathrm{S}\left(\mathsf{J}_x\rho\mathsf{J}_x-\frac{1}{2}\left\{\mathsf{J}_x^2,\rho\right\}\right). \label{eq:SpinPhaseDiffusion}
\end{align}
While most mixed terms vanish because the modes do not overlap spatially or because they are of different parity, we here safely neglect the remaining ones as this only underestimates the influence of CSL. 

The CSL-induced dephasing and diffusion is quantified by the rate parameters 
\begin{subequations}\label{eq:GammaPS}
\begin{align} \label{eq:GammaP}
   \Gamma_{\rm P} &= 2\lambda \frac{m^2}{\rm u^2}
f_\mathrm{P} (r_{C})
\\\label{eq:GammaS}
\Gamma_{\rm S} &= 2\lambda \frac{m^2}{\rm u^2}
f_\mathrm{S} (r_{C})
\end{align}
\end{subequations}
involving $r_{C}$-dependent geometry factors that will be relevant for the mode configurations discussed below. The explicit form of the factors depends on whether we consider strongly elongated condensates or tightly confined ones. 
In the elongated case, the quasi-free direction can be ignored, as it will be integrated out upon readout \cite{Berrada2013,van2014interferometry}. The two interfering modes $a,b$ are then characterized by their transverse two-dimensional wave functions $\langle \boldsymbol{r}|\psi_{a,b} \rangle $, leading to
\begin{align}
    f_\mathrm{P}^{\rm{(1D)}} (r_{C}) &:= \frac{r_{C}^2 }{2\pi} \int \diff^2 {q}\,  e^{-q^2 r_C^2} |W_{aa}(\mathbf{q}) - W_{bb}(\mathbf{q})|^2, \label{eq:fP_2d} \\
    f_\mathrm{S}^{\rm{(1D)}} (r_{C}) &:=\frac{r_{C}^2 }{2\pi} \int \diff^2 {q}\, e^{-q^2 r_C^2} |W_{ab}(\mathbf{q}) + W_{ba}(\mathbf{q})|^2 . \label{eq:fS_2d}
\end{align}
In the tightly confined case, we instead deal with  three-dimensional wave functions characterizing the mode,
\begin{align}
    f_\mathrm{P} (r_{C}) &:= \frac{r_{C}^3 }{2\pi^{3/2}} \int \diff^3 {q}\,  e^{-q^2 r_C^2} |W_{aa}(\mathbf{q}) - W_{bb}(\mathbf{q})|^2, \label{eq:fP_3d} \\
    f_\mathrm{S} (r_{C}) &:=\frac{r_{C}^3 }{2\pi^{3/2}} \int \diff^3 {q}\, e^{-q^2 r_C^2} |W_{ab}(\mathbf{q}) + W_{ba}(\mathbf{q})|^2 . \label{eq:fS_3d}
\end{align}

For spatially separated modes in a typical Mach-Zehnder or double-well configuration, one immediately finds $W_{ab} (\mathbf{q}) \approx 0$. 
The CSL generator \eqref{eq:TwoModeLindblad} then reduces to pure dephasing between the modes, i.e.~a phase noise channel in the collective spin representation \cite{Ma2011},
\begin{align}\label{eq:PhaseDiffusion}
\mathcal{L}\rho &= \Gamma_\mathrm{P}\left(\mathsf{J}_z\rho\mathsf{J}_z-\frac{1}{2}\left\{\mathsf{J}_z^2,\rho\right\}\right), 
\end{align}
To obtain the scaling behavior of the geometry factor $f_\mathrm{P}(r_{C})$ of dephasing rate we consider two identical Gaussian modes at distance $\Delta_x$ much greater that their relevant widths.
For elongated modes this yields
\begin{equation}\label{eq:OverlapDoubleWell}
    f_\mathrm{P}^{\rm{(1D)}} (r_{C}) = \frac{1-\exp\left[-\Delta_x^2/4(w_x^2+r_C^2)\right]}{\sqrt{(1+w_x^2/r_C^2)(1+w_y^2/r_C^2)}},
\end{equation}
and for tightly confined BECs $ f_\mathrm{P} (r_{C}) = f_\mathrm{P}^{\rm{(1D)}} (r_{C}) / \sqrt{1+w_z^2/r_C^2}$.

\begin{figure}
  \centering
  \includegraphics[width=0.39\textwidth]{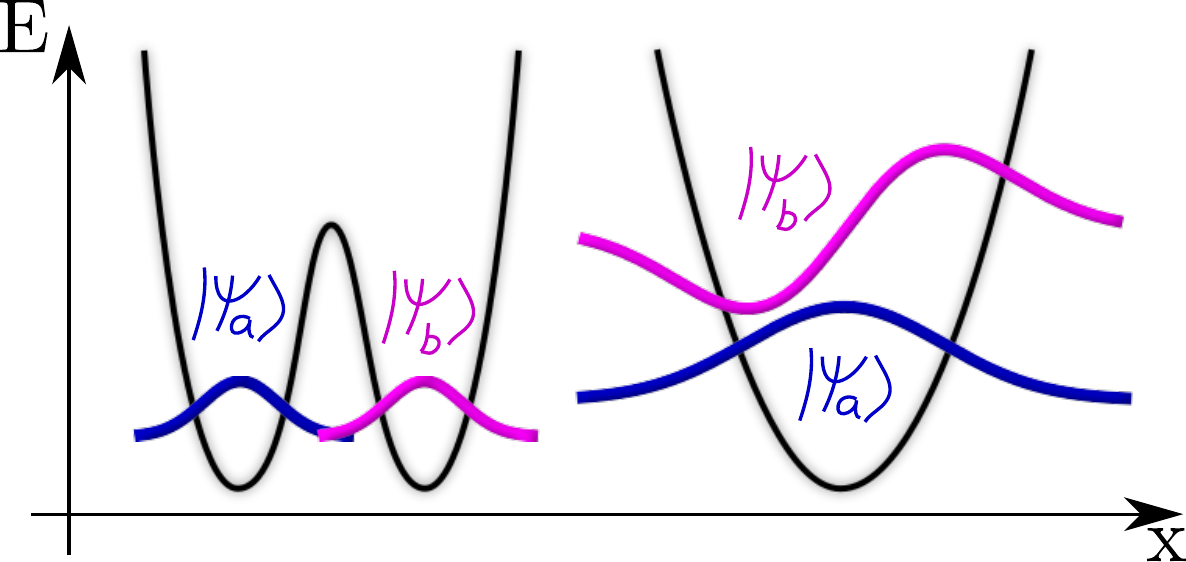}
  \caption{Two typical mode configurations used for Mach-Zehnder interference. Left: In a double-well configuration the two modes can be assumed to have a negligible spatial overlap \cite{Berrada2013}. This description also applies  to free flying modes in an atomic fountain \cite{Kovachy2015,asenbaum2017phase,schrinski2017sensing}. Right: Alternatively, one may operate the Mach-Zehnder scheme with the ground state and the first excited state of a single well configuration \cite{van2014interferometry}. 
  CSL can then lead to transitions between the modes, unlike in the previous case.
    }
\label{fig:Potentiale}
\end{figure}

The situation is more complicated if the two interfering modes overlap spatially, as in a single-well configuration depicted in Fig.~\ref{fig:Potentiale}. Then $W_{ab} $ is finite implying that CSL will not only induce dephasing, but also diffusion in the occupation numbers. For instance, consider a condensate split between the ground state and the first excited state of a harmonic potential in the $x$-direction, with trapping frequency $\omega$ and ground-state width $w_x = \sqrt{\hbar/2m\omega}$. The motion along $y$ and, if applicable, along $z$ shall remain in a harmonic ground state of width $w_y$ and $w_z$, respectively. This leads to
\begin{align}
    W_{aa}^{\rm{(1D)}} (\mathbf{q}) &= e^{-( q_x^2 w_x^2 + q_y^2 w_y^2)/2}, \\
    W_{aa} (\mathbf{q}) &= e^{-( q_x^2 w_x^2 + q_y^2 w_y^2 + q_z^2 w_z^2)/2}, \nonumber
   \end{align}
and, accordingly, $W_{bb} (\mathbf{q}) = \left( 1 - q_x^2 w_x^2 \right) W_{aa} (\mathbf{q})$ and $W_{ab} (\mathbf{q}) = i q_x w_x W_{aa} (\mathbf{q}) = W_{ba} (\mathbf{q})$.
The respective geometry factors are then
\begin{align}\label{eq:OverlapSingleWell}
    f_\mathrm{P}^{\rm{(1D)}} (r_{C}) &= \frac{3 r_{C}^2 w_x^4 }{8\sqrt{r_{C}^2+w_y^2}(r_{C}^2 + w_x^2)^{5/2}}, \nonumber \\
    f_\mathrm{S}^{\rm{(1D)}} (r_{C}) &= \frac{r_{C}^2 w_x^2 }{\sqrt{r_{C}^2+w_y^2}(r_{C}^2 + w_x^2)^{3/2}},
\end{align}
or again $ f_\mathrm{P,S} (r_{C}) = f_\mathrm{P,S}^{\rm{(1D)}} (r_{C}) / \sqrt{1+w_z^2/r_C^2}$.

\section{Combined effect of phase dispersion and diffusion}
\label{sec:dispDiff}

To solve the the master equation  \eqref{eq:LioSup} for condensate time evolution under the CSL-induced dephasing and diffusion \eqref{eq:SpinPhaseDiffusion}, 
we apply the phase-space method of Ref.~\cite{schrinski2019macroscopicity}. 
Consider high atom numbers $N \gg 1$ and coherent superpositions with a well-defined interferometric phase, such that the collective spin representation is sharply localized on the equator of the generalized Bloch sphere, $\langle \mathsf{J}_z \rangle \approx 0$ and $\Delta \mathsf{J}_{x,y,z}^2 \ll N^2$. To a very good approximation, the state is then described by a continuous Wigner function $w_t(\varphi,n)$ in the flat tangential phase space of the conjugate variables $n$ (population difference) and $\varphi$ (interferometric phase angle), provided the support of $w_t$ is limited to $|n| \ll N$ and stretches over an angular region much less than $2\pi$.
(The latter constraint can be alleviated with help of a trigonometric mapping of the Wigner function onto a periodic function over the equator, $\varphi \in [0,2\pi)$, but this will not be necessary for the cases studied here.) 
In a rotating frame that absorbs the free linear evolution of the phase, $\varphi (t) = \varphi(0) + \epsilon t/\hbar$, the master equation \eqref{eq:LioSup} translates into the Fokker-Planck equation
\begin{align}\label{eq:WignerFunction}
\partial_t w_t(\varphi,n) \approx&-\zeta n\partial_\varphi w_t(\varphi,n)+\frac{N^2\Gamma_\mathrm{S}}{4}\partial_n^2w_t(\varphi,n)\nonumber\\
&+\frac{\Gamma_\mathrm{P}}{2}\partial_\varphi^2w_t(\varphi,n).
\end{align}
Here, a $\varphi$-dependence in the second term was approximated by an angular average, which is permissible for small diffusion rates, $\Gamma_{\rm S} \ll \epsilon/\hbar$, provided
the interference time extends over at least one free oscillation period.

\begin{figure*}
  \centering
  \includegraphics[width=0.89\textwidth]{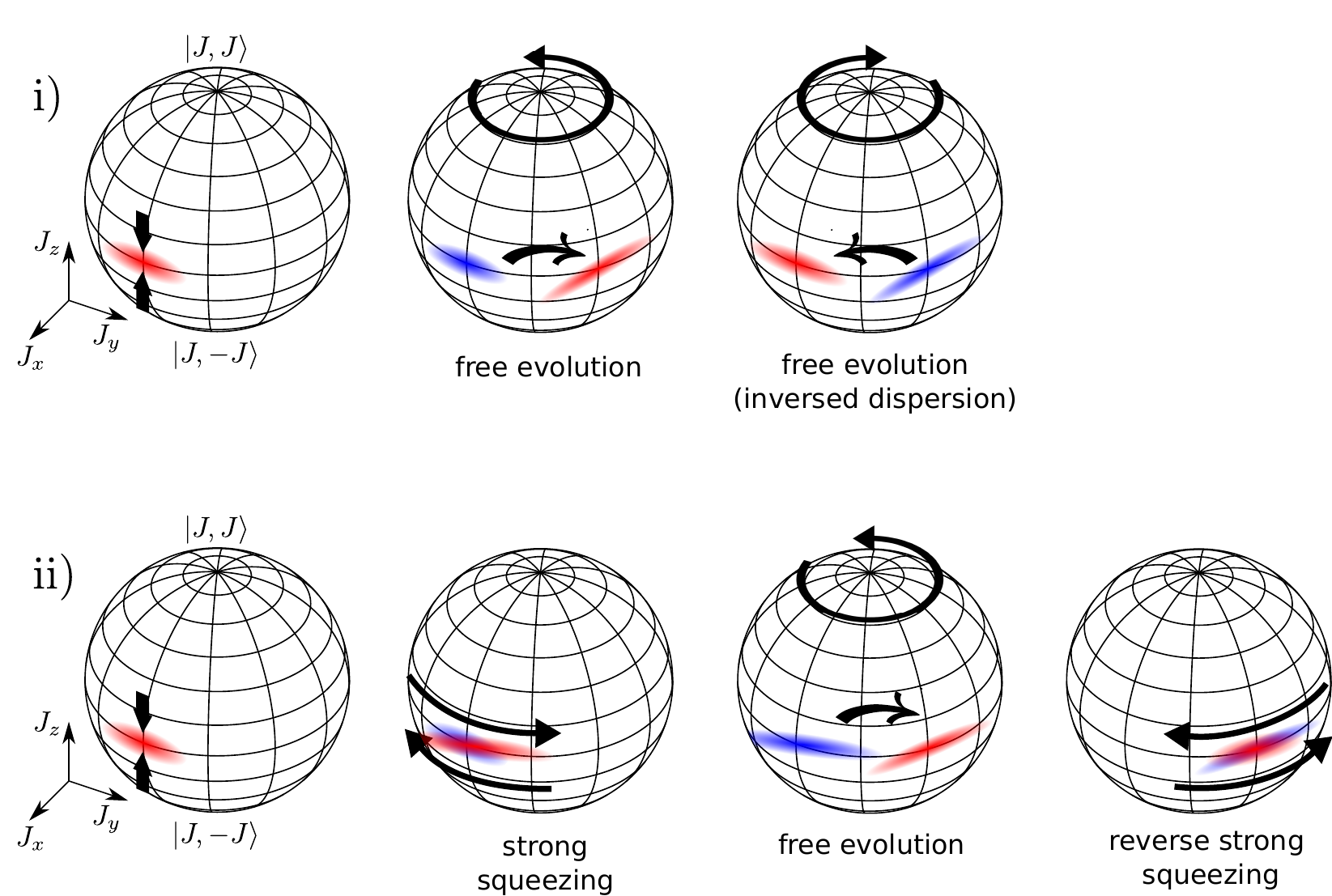}
  \caption{Proposed alternative interference protocols in which the coherent phase broadening caused by dispersion is undone by tuning the two-body interaction strength. In the two-step protocol (i), the interaction is flipped from attractive to repulsive (or vice versa) after half the interference time to undo the dispersion. Any observed phase broadening then indicates the presence of dephasing and diffusion, where the latter is enhanced by the dispersion effect. The dispersion enhancement of CSL diffusion can be improved in the three-step protocol (ii) by amplifying (and undoing) the dispersion for a short time before and after a long stretch of free evolution under weak dispersion. Both protocols benefit from an initially number-squeezed condensate state, but (ii) features an additional 12-fold improved sensitivity on the CSL rate  compared to (i). }
\label{fig:EchoSchemes}
\end{figure*}

Equation \eqref{eq:WignerFunction} constitutes a Gaussian channel, and every initially Gaussian Wigner function will thus remain Gaussian. In particular, the $\varphi$-marginal of the Wigner function, which represents the phase distribution of the interfering state after time $t$, will be of the form 
\begin{align}
\int dn\,w_t(\varphi,n)=\frac{1}{\sqrt{2\pi}\sigma_\varphi(t)}\exp\left[-\frac{\varphi^2}{2\sigma_\varphi^2(t)}\right].
\end{align}
More generally, we can obtain an exact solution for arbitrary initial states with help of the characteristic function in Fourier space,
\begin{equation}\label{eq:chit_differential}
    \chi_t (s,q) := \int \diff \varphi \diff n\, w_t (\varphi,n) e^{is \varphi + iq n}.
\end{equation}
It evolves according to
\begin{equation}
    \partial_t \chi_t (s,q) = \zeta s \partial_q \chi_t (s,q) - \left( \frac{\Gamma_{\rm P}}{2} s^2 + \frac{N^2 \Gamma_{\rm S}}{4} q^2 \right) \chi_t (s,q),
\end{equation}
which yields the solution
\begin{align} \label{eq:chit_sol}
    \chi_t (s,q) &= \chi_0 (s,q +\zeta t s) \\
    &\times \exp \left[ - \frac{\Gamma_{\rm P}t}{2} s^2 - \frac{N^2 \Gamma_{\rm S}t}{4} \left( q^2 + \zeta t q s + \frac{\zeta^2 t^2}{3} s^2 \right) \right]. \nonumber
\end{align}
The characteristic function generates the moments of the phase distribution as $\langle \varphi^k \rangle_t = (-i\partial_s)^k \chi_t(0,0)$. Plugging the above solution into $\sigma_\varphi^2 (t) = \langle \varphi^2 \rangle_t - \langle \varphi \rangle_t^2$ leads to 
\begin{align}\label{eq:phaseVar_withDiff}
\sigma_\varphi^2(t)=\sigma_\varphi^2(0)+\Gamma_\mathrm{P} t+\zeta^2t^2\left(\sigma_n^2(0)+\frac{\Gamma_\mathrm{S}N^2t}{6}\right),
\end{align}
provided that initially $\langle n(0)\rangle=0$ and $ \langle n(0)\varphi(0) \rangle = 0$. By carrying out many repetitions of the standard interference scheme depicted in Fig.~\ref{fig:RamseyScheme} with $N\gg 1$ atoms per run, one samples a Gaussian phase distribution with variance $\sigma_\varphi^2$.

Equation \eqref{eq:phaseVar_withDiff} indicates that the sensitivity of the standard interference scheme to CSL effects may be severely impacted by the presence of atom-atom interactions. Specifically, the first term in brackets may dominate the broadening of the final distribution for finite interaction-induced dispersion $\zeta$. 
We therefore consider two modified interference schemes in the following, which serve to suppress the broadening, see Fig.~\ref{fig:EchoSchemes}. 

The first scheme, depicted in Fig.~\ref{fig:EchoSchemes} (i), cancels the regular broadening by reversing the direction of dispersion after half the interference time. This may be achieved by switching from an attractive to an equally strong repulsive atom-atom interaction or vice versa. The corresponding condensate state can be calculated by first computing $\chi_{t/2} (s,q)$ according to \eqref{eq:chit_sol} and then inserting the result as the new initial condition into \eqref{eq:chit_sol} with $\zeta$ replaced by $-\zeta$, 
\begin{align}\label{eq:chit_sol_echo}
    \chi_t (s,q) &= \chi_0 (s,q) \exp \left[- \frac{\Gamma_{\rm P}t}{2} s^2 - \frac{N^2 \Gamma_{\rm S}t}{4} \left( q^2 + \frac{\zeta^2 t^2}{12} s^2 \right) \right]. 
\end{align}
This results in the detectable phase variance
\begin{align} \label{eq:phaseVar_echoScheme}
    \sigma_\varphi^2 (t) &= \sigma_\varphi^2 (0) + \Gamma_{\rm P}t + \frac{N^2 \Gamma_{\rm S}t \zeta^2 t^2 }{24},
\end{align}
which has the $\sigma_n^2(0)$-term removed as desired, at the price of a mere four-fold reduction of the CSL diffusion broadening. Needless to say, an experimental realization requires an exact and precisely timed flipping of the sign of $\zeta$; practical limitations due to three-body losses will be discussed in Sec.~\ref{Statistics}.

Alternatively, one may consider the three-step protocol depicted in Fig.~\ref{fig:EchoSchemes}(ii), which separates the regular dispersion and the CSL-induced diffusion effect in time. Starting from an unsqueezed or squeezed superposition state, we first switch on a strong dispersion $\zeta_1$ for a short time $t_1 \ll t$ and allow the condensate to shear across the equator of the Bloch sphere. Then we switch off the dispersion by tuning the atom interactions to $|\zeta_2|\ll|\zeta_1|$ and let the condensate evolve under CSL for the largest part of the interrogation time, $t_2 = t-2t_1$. Finally, we revert the shearing ($\zeta_3t_3 = -\zeta_1t_1-\zeta_2t_2$) to undo the accumulated dispersion during another short period $t_3 = t_1$. 
As before, in the absence of CSL-induced dephasing and diffusion, the protocol would map the condensate state back onto the initial state on the co-rotating Bloch sphere. Under CSL, however, the final condensate state becomes
\begin{align}\label{eq:chit_sol_3steps}
    \chi_t (s,q) &= \chi_0 (s,q) \exp \left[- \alpha s^2 - \beta qs -\gamma q^2 \right],
\end{align}
with
\begin{align}
\alpha=&\frac{\Gamma_\mathrm{P}t}{2}+\frac{N^2\Gamma_S }{12}[\zeta_1^2t_1^2(2t_1+3t_2)+4\zeta_2^2t_2^3+9\zeta_1\zeta_2t_1t_2^2],   \nonumber\\ 
\beta=&-\frac{N^2\Gamma_S }{4}t_2(\zeta_2t_2+2\zeta_1t_1)   , \nonumber\\
\gamma=&\frac{N^2\Gamma_\mathrm{S} t}{4}.    
\end{align}
This results in the phase broadening
\begin{align} \label{eq:phaseVar_3steps}
    \sigma_\varphi^2 (t) = \sigma_\varphi^2 (0) + 2\alpha 
    \geq \sigma_\varphi^2 (0) + \Gamma_{\rm P}t + \frac{N^2 \Gamma_{\rm S} t_2 \zeta_1^2 t_1^2 }{2},
\end{align}
where in the last inequality we only include the leading order contribution for $t_2\gg t_1$ and $\zeta_1 t_1\gg \zeta_2 t_2$.  
Comparing the phase variances \eqref{eq:phaseVar_echoScheme} and \eqref{eq:phaseVar_3steps} of the two protocols, we see that despite $t_1 \ll t_2\simeq t$, a suitably amplified dispersion $\zeta_1$ can still lead to a stronger broadening in the three step protocol, viz.~$\sqrt{12}\zeta_1 t_1 > \zeta t$. In particular, equating $\zeta_1 t_1 = \zeta t$ for comparison leaves us with a 12-fold greater impact of CSL diffusion in protocol (ii) compared to (i).

Note that, in principle, a finite dispersion $\zeta$ can also lead to (partial) revivals of the initially prepared condensate state at (fractions of) the revival time $t_{\rm rev} =2\pi/\zeta$. Hence, as an alternative to the discussed multi-step protocols, the direct observation of those revivals at fixed $\zeta$ would be highly sensitive to CSL-induced diffusion broadening as well. However, this would require long interrogation times, barely feasible in practice, and the predictions would not be covered by the continuous phase-space approximation of the condensate state.

\section{Scaling with particle number}\label{ScalingProp}

In the previous section, we quantified the broadening of the condensate phase distribution over time due to CSL-induced dephasing and diffusion, comparing three different interferometer protocols. 
The respective phase variances \eqref{eq:phaseVar_withDiff}, \eqref{eq:phaseVar_echoScheme}, and \eqref{eq:phaseVar_3steps}, together with the expressions for the CSL rates  \eqref{eq:GammaPS} 
in terms of the associated geometry factors $f_{\rm{P,S}} (r_C)$ for the chosen interference modes, fully characterize the CSL effect as a function of its parameters $\lambda$ and $r_C$. Inferring the phase variance from a measured sample of atom count differences of the recombined condensate, one can proceed to estimate the range of CSL parameters compatible with or excluded by the data.

Before heading to the details of noisy measurement statistics, it is worthwhile to study how CSL sensitivity scales with the experimental parameters, specifically the atom number $N$, and how that compares to other proposed interferometric CSL tests. Notably, in the introduction we made the point that regular low-contrast interference of a Bose-Einstein-condensed atom cloud is just as sensitive against the CSL effect as the consecutive interference of the same number of single atoms. This fundamentally limits its prospects as a CSL test compared to molecule or NOON-state interferometry where the sensitivity scales with the square of the atom number due to the high degree of entanglement \cite{bilardello2017collapse}. Optomechanical CSL tests, on the other hand, exhibit an at most linear scaling of CSL sensitivity with mass (or atom number) \cite{schrinski2022macroscopic}, but they are much more massive to begin with. 

Our main point here is that precision measurements of interfering two-mode BECs close to the shot noise limit can lead to linear and even superlinear scaling of CSL sensitivity with the atom number $N$. In fact, the linear scaling can already be seen for two spatially separated modes, which are not affected by CSL-induced diffusion,  $\Gamma_\mathrm{S}=0$. Interfering them according to the standard Mach-Zehnder protocol, the phase variance \eqref{eq:phaseVar_withDiff} after the interrogation time $t$ simplifies to
\begin{align}\label{eq:phaseVar_dephasing}
\sigma_\varphi^2(t)=\sigma_\varphi^2(0)+\Gamma_\mathrm{P} t+\zeta^2t^2\sigma_n^2(0).
\end{align} 
Now consider an initial condensate state close to the minimum of the number-phase uncertainty relation, $\sigma_{\varphi}^2(0) \sigma_n^2 (0) =1$, which may achieve near-unit interference visibility. This could be a product state of $N$ two-mode superpositions, with shot noise limited $\sigma_\varphi^2 (0) =1/ \sigma_n^{2} (0) = 1/N$, or a phase-squeezed state with  $\sigma_\varphi^2(0) = \xi_0^2/N$ and $\xi_0 < 1$ below shot noise \cite{Ma2011} \footnote{We remark that the minimum uncertainty product $\sigma_{\varphi}^2(0) \sigma_n^2 (0) =1$ can not be attained if phase squeezing is generated through one-axis twisting \cite{Kitagawa1993}}. For a non-interacting condensate, $\zeta=0$, this implies that the condensate could resolve phase spreads $\Gamma_{\rm P}t$ caused by CSL (or any other dephasing channel) as small as $\xi_0^2/N$.

In the presence of dispersion, the (anti-squeezed) conjugate number variance $\sigma_n^2 (0)$ may cause a substantial phase broadening in  \eqref{eq:phaseVar_dephasing} 
which is detrimental to the interference visibility. Indeed, measuring a large phase variance after recombination, $\sigma_\varphi^2(t)\sim1\gg\xi_0^2/N$, renders the initial $1/N$ resolution irrelevant, see Fig.~\ref{fig:Gaussians}. This amounts to the regime of low interference contrast in experiments, typically due to uncontrolled phase noise and other sources of error. 

To clarify how CSL sensitivity scales with $N$, suppose we infer an effective squeezing $\xi_t^2 = N \sigma_\varphi^2 (t) $ from a measured sample at known initial state parameters $(N,\xi_0,\sigma_n)$ and dispersion $\zeta$. Then Eq.~\eqref{eq:phaseVar_dephasing} implies that the data are consistent with CSL rate parameters
\begin{align}\label{eq:shotNoiseScalingPhaseFlips}
\lambda \leq \frac{({\rm u}/m)^2}{2 N t } \frac{\xi_t^2 - \xi_0^2 - \zeta^2 t^2 N \sigma_n^2 (0)  }{f_\mathrm{P}(r_C)},
\end{align}
and greater rate parameters are ruled out.
The scaling with $1/Nt$ highlights the trade-off between measurement resolution and interference time: a short-time precision measurement with a minimum-uncertainty condensate of many atoms, ideally with as low $\zeta$ as possible \footnote{In the presence of phase dispersion one could, in principle, subtract the  effect from the data if $\zeta \neq 0$ is known precisely. In practice, however, more data would then be needed to resolve a small broadening on top of the dispersion effect, as discussed in Sect.~\ref{Statistics}.}, can be on par with a conventional long-time interferometer operating a small condensate or individual atoms.
Still, the linear growth of CSL sensitivity with the atom number falls short of the quadratic mass scaling offered by interferometry with rigid compounds. We show next that, in a different mode configuration, BEC interferometry does offer an equivalent $N^2$ scaling.

The key idea is to consider two interfering modes with a significant spatial overlap, instead of separated arms, as shown in Fig.~\ref{fig:Potentiale}. 
The interplay between interaction-induced phase dispersion and CSL-induced diffusion results in an $N^2$-amplified overall impact of CSL on the phase distribution of the two-mode state:  The inter-atomic interactions cause the energy splitting between the modes to depend on the number difference $n$. The ensuing dispersion leads to a phase spread that grows with the conjugate atom number uncertainty, which in turn increases by virtue of the CSL-induced  diffusion effect, see Eq.~\eqref{eq:phaseVar_withDiff}.
In a measurement with an interacting BEC at sufficiently large $N$ and $\zeta$, the diffusion term quickly exceeds the dephasing contribution $\Gamma_{\rm P} t$, which results in an improved CSL bound. Assuming, as before, that one infers an effective $\xi_t$ from a sample and conservatively attributes all incoherent broadening to CSL, the data would be consistent with
\begin{align}\label{eq:shotNoiseScalingSpinFlips}
\lambda &\leq \frac{({\rm u}/m)^2}{2N t} \frac{\xi_t^2-\xi_0^2-\zeta^2t^2 N \sigma_n^2 (0)}{f_\mathrm{P}(r_C) + (N^2/6) \zeta^2 t^2 f_\mathrm{S}(r_C)} \nonumber \\
& < 3 \frac{({\rm u}/m)^2}{N^3t^3 \zeta^2}\, \frac{\xi_t^2-\xi_0^2-\zeta^2t^2 N \sigma_n^2 (0)}{f_\mathrm{S}(r_C)}.
\end{align}
In the second line, we omit the dephasing term, which is typically negligible in overlapping interacting condensates with $N\gg 1$. This only underestimates the  sensitivity to CSL falsification.

Equation \eqref{eq:shotNoiseScalingSpinFlips} shows that to rule out small CSL rates, one would have to detect a diffusion-induced broadening \emph{on top of} a potentially large systematic broadening caused by dispersion alone, which requires precise knowledge of $\sigma_n(0)$ and more measurement data (see below). 
To alleviate this problem, we propose two echo-like interference protocols in the previous section, see Fig.~\ref{fig:EchoSchemes}, in which the dispersion broadening cancels. 

The two protocols (i) and (ii) demand varying levels of experimental control over the atom-atom interaction, e.g.\ via a Feshbach resonance \cite{Gustavsson2008,fattori2008magnetic}: 
protocol (i) requires a single control step that switches from attractive to repulsive interaction ($\zeta \to -\zeta$) after half the interference time, while (ii) requires two control steps switching from a strong to a weak and then to a strong dispersion of opposite sign. We assume that the switching can be performed on a much shorter time scale than the interference time $t$. Omitting the dephasing term as in \eqref{eq:shotNoiseScalingPhaseFlips}, the diffusion phase broadenings predicted in \eqref{eq:phaseVar_echoScheme} and \eqref{eq:phaseVar_3steps} lead to the CSL bounds
\begin{align}\label{eq:shotNoiseScalingSpinFlipsWithoutDispersion}
\lambda^{\rm (i)} &< 12 \frac{({\rm u}/m)^2}{N^3t^3 \zeta^2}\, \frac{\xi_t^2-\xi_0^2}{f_\mathrm{S}(r_C)}
\intertext{and}
\lambda^{\rm (ii)} &< \frac{({\rm u}/m)^2}{N^3 t \zeta_1^2 t_1^2}\, \frac{\xi_t^2-\xi_0^2}{f_\mathrm{S}(r_C)}
\end{align} 
for the two dispersion-compensating protocols. For the three-step case (ii), we assume in addition that $t_1 \ll t_2 \approx t$. The resulting CSL bound (at $\zeta_1 t_1 = \zeta t$) resembles that of (i), but without the prefactor 12---an improvement by just over one order of magnitude.

Equations \eqref{eq:shotNoiseScalingSpinFlips} and \eqref{eq:shotNoiseScalingSpinFlipsWithoutDispersion} demonstrate the key point of our proposal: interferometric tests of CSL with atom condensates can exhibit a favourable scaling with the atom number $N$ that is on par with proposed matter-wave experiments, which envisage interfering massive molecules or nanoparticles that contain a similar number of constituent atoms.
While the atom-atom interaction may lead to a transient buildup of correlations in the condensate, our scheme neither requires the detection of many-atom correlations nor the preparation of highly entangled cat or NOON states. In fact, the $N$-scaling applies also to initially uncorrelated condensates ($\xi_0=1$). 
Compare this to an equivalent two-arm interferometer with individual nanoparticles of, say, $N$ identical constituent atoms, which measures an interference visibility $\mathcal{V}$ after the interference time $t$. The associated CSL bound, $\lambda \leq ({\rm u}/M)^2 |\ln \mathcal{V}|/f_{\rm P} (r_C) t$, then improves with $N$ via the particle mass $M = N m$, provided that the particle is much smaller than $r_C$.

In the next section, we will corroborate our  assessment of the CSL sensitivity scaling with the number $N$ of Bose-Einstein-condensed particles by accounting for the most relevant experimental limitations including finite measurement data.

\section{Measurement statistics and experimental effort}\label{Statistics}

\begin{table*}
\begin{center}
\caption{\label{tab:parameters} Parameters and achievable sensitivities for CSL tests in single-well (SWI) and double-well (DWI) MZI setups with $^{87}$Rb and $^{174}$Yb atoms at different interference times $t$. For a given atom number $N$, ground-state width $w_x$, and phase squeezing parameter $\xi_0$, we list the (rounded) number of measurement runs $k$ needed to exclude CSL rates greater than $\lambda_\mathrm{min}$ at $25\,\%$ precision. The phase variance due to two-body interactions  is denoted as $\sigma_{\rm 2b}^2$, while $\sigma_{\rm 3b}^2$ gives the phase spread due to three-body recombination.
For the setup (i), phase dispersion is inverted halfway to cancel its influence on phase broadening, whereas for (ii), we assume strong dispersion (achieving $\zeta\tau\simeq0.00157$) over a short period $\tau \ll t$ before and after the interference to further magnify the CSL-induced diffusion effect, as detailed in the main text and depicted in Fig.~\ref{fig:EchoSchemes}. In both cases (i) and (ii) the condensate is in the three-dimensional ground state to ensure that dispersion can be reversed without coherence loss \cite{widera2008quantum}. The scattering length is assumed to be reduced to $1\%$ (Rb DWI), or $10\%$ (Rb SWI and Yb $\mathrm{DWI}^{(i)}$); we do not assume that the three-body interaction can be manipulated significantly. The DWI arm separation is taken as $\Delta_x=10 w_x$, and in the single-well configuration $w_y$ (and $w_z$ if the third dimension is not traced out) is chosen as $\sqrt{0.7}w_x$ to lift degeneracy. The respective exclusion curves of all four setups are depicted in Fig.~\ref{fig:CSL_exclusion}}
 \begin{ruledtabular}
 \begin{tabular}{l c c c c } 
 
 setup & Rb DWI & Rb SWI & Yb DWI$^{(\rm i)}$ & Yb SWI$^{(\rm ii)}$\\
 \hline
 Atom number $N$ & $3\times10^4$ & $2.4\times10^6$ & $1\times 10^4$ & $ 5\times 10^5$\\
 Condensate width $w_x$ & $6\,\mu$m&$6\,\mu$m &$4\,\mu$m & $4\,\mu$m\\
 Condensate width $w_y$ & $6\,\mu$m&$5\,\mu$m &$4\,\mu$m & $3.3\,\mu$m\\
 Condensate width $w_z$ & $60\,\mu$m&$60\,\mu$m &$4\,\mu$m & $3.3\,\mu$m\\
 Number density & $1.4\times10^{13}\,\mathrm{cm}^{-3}$ & $7.4\times10^{14}\,\mathrm{cm}^{-3}$ & $1.6\times10^{14}\,\mathrm{cm}^{-3}$ & $7.8\times10^{15}\,\mathrm{cm}^{-3}$\\
 Scattering length & $0.8\,$\AA & $8\,$\AA & $5.5\,$\AA & $55\,$\AA\\
  Initial squeezing $\xi_0^2$ & 0 dB &-20 dB & 20 dB& 30 dB\\
Interference time $t$ & $100\,\mathrm{ms}$& $1\,\mathrm{s}$&$100\,\mathrm{ms}$ & $0.1\,\mathrm{ms}$\\
   Dispersion rate $\zeta$ or $\zeta_2$ & $3.4\times 10^{-4}$ Hz & $4\times 10^{-3}$ Hz & $1.2\times 10^{-2}$ Hz & 0.57 Hz \\
   Strong dispersion $(\zeta_1,t_1)$ & --- & --- & --- & $(56\,\mathrm{kHz},1\,\mu\mathrm{s})$ \\   
  Two-body spread $\sigma_{\rm 2b}^2$ & $3.4\times 10^{-5}$&$0.39$ &  reversed& reversed\\
   Three-body loss rate $\gamma_{\rm 3b}$ & 1 mHz & 10 Hz & 0.1 Hz & 500 Hz \\
Three-body spread $\sigma_{\rm 3b}^2$ &$3.7\times10^{-9}$ & $0.012$& $2\times10^{-6}$& $1\times10^{-7}$\\
\hline
CSL rate $\lambda_\mathrm{min}$ & $3.6\times 10^{-10}$\,Hz & $1.8\times10^{-14}$\,Hz & $2.5\times10^{-12}$\,Hz & $1.2\times 10^{-18}$\,Hz \\
Localization scale $r_C (\lambda_{\min})$ & $8.0\times10^{-6}$m& $5.7\times10^{-6}$m & $1.1\times10^{-5}$m & $4.5\times 10^{-6}$m\\
Measurement repetitions    $k$ & $7.2\times10^5$& $4.4\times10^5$ & $9.2\times10^5$& $9.9\times10^5$\\
 \end{tabular}
\end{ruledtabular}
 \end{center}
\end{table*}

When proposing experimentally challenging tests of ever weaker collapse models, it is not enough to ensure that the proposed setup parameters are within reach of current or future technology. One must also assess realistically the minimum measurement effort and time for a conclusive outcome, in the presence of unavoidable losses and decoherence effects.

While in the early years of cold atom technology, BEC generation was limited in time primarily by low atomic densities and correspondingly slow evaporative cooling rates, steady advances have nowadays enabled BEC production rates of more than 1 Hz \cite{rudolph2015high,vuletic1998degenerate, hu2017creation}.
This goes hand in hand with all-optical cooling to quantum degeneracy \cite{Schreck2021,hu2017creation,Stellmer2013}, which does not suffer from unavoidably strong atom loss during evaporative cooling. 
In Table \ref{tab:parameters}, we therefore assume that the time for a single experimental run is mainly limited by the interference time, which allows for fast duty cycles.  

The relevant figure of merit is the number of measurement repetitions $k$ and the corresponding total integration time $k t$, which for a viable proposal must not be unreasonably long. The minimum $k$ is given by the number of sample points one needs in order to determine the width of the atom count distribution and extract a lower bound of falsified CSL rates $\lambda$  at the desired precision. We can estimate $k$ with help of the Cram\'{e}r-Rao bound, or Bernstein-von Mises theorem \cite{von2014bayesian}: In the presence of CSL, theory predicts a probability distribution of atom count differences $p(n|\lambda; r_C, I)$ conditioned on the CSL rate $\lambda$ at a given CSL length $r_C$ and background information $I$ (which subsumes all relevant experimental parameters, including $N$, $t$, $\bar\varphi$). The corresponding Fisher Information (FI) \cite{fisher1990statistical} then bounds the precision of the $\lambda$ estimate from the data by $\Delta \lambda \geq 1/\sqrt{k \mathcal{I}(\lambda|r_C,I)}$, in the limit of large $k$.

The outcome distributions $p(n|\lambda; r_C, I)$ for the proposed interference schemes are very well approximated by Gaussians, and Eqs.~\eqref{eq:shotNoiseScalingPhaseFlips}, \eqref{eq:shotNoiseScalingSpinFlips}, and \eqref{eq:shotNoiseScalingSpinFlipsWithoutDispersion} define the required consistent (unbiased) $\lambda$-estimators as linear functions of the estimated phase variances $\sigma^2_\varphi (t) = \xi^2_t/N$. If we now separate the CSL contribution proportional to $\lambda$ from the other terms in this expression, $\sigma^2_\varphi (t) = \sigma^2_{\rm conv}(t) + \alpha^2_{\rm CSL}(t) \lambda$, the FI and the respective Cram\'{e}r-Rao bound can be written explicitly as 
\begin{align}
    \mathcal{I}(\lambda|r_C,I) &= \frac{1}{2[\sigma^2_{\rm conv}(t)/\alpha^2_{\rm CSL}(t) + \lambda]^2} 
\end{align}
and
\begin{align}\label{eq:delta}
\frac{\Delta \lambda}{\lambda} & \geq \sqrt{\frac{2}{k}} \left[ 1 + \frac{\sigma^2_{\rm conv}(t)}{\lambda \alpha^2_{\rm CSL}(t)} \right] > \sqrt{\frac{2}{k}} \frac{\sigma^2_{\rm conv}(t)}{\lambda \alpha^2_{\rm CSL}(t)}.  
\end{align}
Hence a CSL test at a \emph{fixed} relative uncertainty $ \delta:=\Delta \lambda/\lambda$ requires at least
\begin{equation}\label{eq:CramerRao_k}
k \geq \frac{2}{\delta^2} \left[ 1 + \frac{\sigma^2_{\rm conv}(t)}{\lambda \alpha^2_{\rm CSL}(t)} \right]^2 > \frac{2}{\delta^2} \frac{\sigma^4_{\rm conv}(t)}{\lambda^2 \alpha^4_{\rm CSL}(t)}
\end{equation}
measurement repetitions. The second inequalities in \eqref{eq:delta} and \eqref{eq:CramerRao_k}  yield approximate bounds for probing the relevant  regime of small $\lambda$-values at which the CSL-induced broadening is dominated by the conventional phase spread. 

Let us now illustrate how the scaling of CSL sensitivity with the particle number $N$ described in the previous section is reflected in the expected measurement statistics. In the simplest case of a non-interacting condensate and two spatially separated arms (e.g.\ of a double-well potential), there is neither dispersion nor CSL diffusion and the final phase variance \eqref{eq:phaseVar_withDiff} splits into $\lambda \alpha^2_{\rm CSL}(t) = \Gamma_{\rm P}t$ and 
$\sigma_{\rm conv}^2(t) = \sigma_{\varphi}^2 (0) = \xi_0^2/N$. While the CSL term does not directly depend on $N$, the initial phase variance is inversely proportional and thus reduces the required number $k$ of measurement runs with $1/N^2$. If one instead interfered \textit{single} atoms in an equivalent number of $kN$ independent  repetitions, each atom would be subject to an independent dephasing channel. Hence the precision of the $\lambda$-estimate would scale less favourably like $\delta \propto 1/\sqrt{kN}$.

In comparison, the presented two- and three-step interference schemes featuring interacting condensates in spatially overlapping modes by far surpass single-atom measurements. Taking only the leading CSL diffusion terms in \eqref{eq:phaseVar_echoScheme} and \eqref{eq:phaseVar_3steps} into account, we have $\lambda \alpha^2_{\rm CSL}(t) \propto N^2 \Gamma_{\rm S} t$.

Conversely, any additional known source of decoherence or phase noise in the experiment will rapidly increase $\sigma_{\rm conv}^2 (t)$ and thus $k$, calling for precision measurements close to the phase uncertainty limit. 
Most common noise sources are due to technical limitations (stability of trap lasers, collisions with residual background gas etc.) that are likely to be improved in the future. Other limitations of two-mode BEC states are rather fundamental and hence unavoidable:
With a growing number $N$ of atoms condensed into a mode volume $V$, two- and three-body scattering and recombination processes compromise the phase and number stability of the BEC. 

The dominant effect of two-body collisions is phase dispersion, as quantified in terms of the rate $\zeta$ in Eqs.~ \eqref{eq:shotNoiseScalingPhaseFlips} and \eqref{eq:shotNoiseScalingSpinFlips}. Given the s-wave scattering length $a$, we have $\zeta = 4\pi\hbar a / m V$
for an evenly split two-mode condensate in linear approximation \cite{Javanainen1997PhaseDispersion,Berrada2013}. Depending on the chosen atom species and volume, $a$ must be Feshbach-tuned \cite{Gustavsson2008,fattori2008magnetic} in order to suppress (or amplify) the associated phase broadening compared to CSL diffusion. 

The main source of condensate depletion at higher atom densities is three-body recombination, causing a loss rate $\gamma_{\rm 3b} \approx K_{\rm 3b} (N/V)^2$ per atom. Note that the characteristic rate constant $K_{\rm 3b}$ remains finite even at vanishing scattering lengths due to the influence of low-energy Efimov resonances \cite{gross2009observation}. While the authors of \cite{bilardello2017collapse} assumed that $K_{\rm 3b}$ could be suppressed further, we take the associated three-body loss as a constraint that most probably cannot be overcome by future technological improvements, limiting the accessible range of CSL lengths $r_C$ at feasible condensate densities. (At large $a$, the loss amplifies like $K_{\rm 3b} \propto a^4$.) 
Among the lowest known $K_3$ values are those of Rubidium ($5.8\times10^{-30}\,\mathrm{cm}^6$/s) \cite{burt1997coherence} and of Ytterbium ($4\times10^{-30}\,\mathrm{cm}^6$/s) \cite{takasu2003spin}, which will be our prime candidates. Regarding CSL sensitivity, the relevant effect of atom loss is the effective phase broadening it implies. Given that $N_{\rm 3b} (t) \ll N$ atoms are lost, we can estimate a small additional phase spread of $\sigma_{\rm 3b}^2 (t) = N_{\rm 3b}(t)/[N-N_{\rm 3b}(t)]N \approx N_{\rm 3b} (t)/N^2$ \cite{Ma2011} that must be added to $\sigma^2_{\rm conv}(t)$ in Eq.~\eqref{eq:CramerRao_k}. The small reduction of $N$, or any small fluctuation of the atom number for that matter, can always be addressed by using conditioned likelihoods in the data analysis \cite{schrinski2019macroscopicity}.

Taking the aforementioned fundamental constraints on measurement time and condensate density into account, we propose exemplary interferometer setups that shift the boundaries of interferometric CSL tests far beyond the current status quo.
Table \ref{tab:parameters} lists four experimental schemes including their key parameters and compares the required number of measurement repetitions $k$ to test CSL rates $\lambda \geq \lambda_{\min}$ with precision $\delta = 0.25$, according to the Cram\'{e}r-Rao bound. 
While precision measurements in the double-well configuration may improve existing interferometric CSL tests by a few orders of magnitudes, they will most likely not improve the current bounds provided by non-interferometric tests, much less probe the originally proposed CSL rate $\lambda=10^{-16}$Hz at $r_C=10^{-7}$m. 
The key constraint is the three-body recombination that cannot be suppressed by decreasing the scattering length $a$. For this reason, it is most favorable to reduce the atom density by employing large trap volumes; the shallowest state-of-the-art traps can reach frequencies as low as $\omega\sim 10\,\mathrm{Hz}$ \cite{van2010bose,muntinga2013interferometry}. This requires a preparation time of the condensate $t_\mathrm{prep}\gtrsim 10/\omega$, and we ensured that the total run time of the proposed experiment $k(t+t_\mathrm{prep})$ does not exceed $10^6$\,s, a typical value considered in experimental proposals aimed at fundamental tests with BECs \cite{dimopoulos2007}.

On the other hand, for single-well interferometry, the combined effect of CSL-induced diffusion and dispersion due to two-body interactions can outscale the three-body recombination rate. It is in principle possible to rule out CSL rates as low as the originally proposed value of $10^{-16}\,$Hz with our proposed three-step interference protocol, provided that one can reliably prepare large number-squeezed condensates and tune their dispersion on short time scales. For example, the best exclusion curve in Fig.~\ref{fig:CSL_exclusion} would be achieved with the parameters in the right column of Table I: a rather dense condensate of $N=5\times10^5$ ytterbium atoms, initialized in a strongly number-squeezed state and interfered over $100\,\mu$s, with additional $1\,\mu$s-intervals of strongly boosted (and sign-flipped) dispersion by $5$ orders of magnitude. 
This experimental scheme may be still far from the current state of the art, but it should be fundamentally feasible.

\section{Generalizations of the CSL model} \label{Generalizations of the CSL model}

So far, we  considered the widely studied original CSL model. More recently, modified models have been suggested that would mitigate CSL-induced energy excitation effects by either introducing thermalization or colored noise. We will now only qualitatively discuss why the here proposed setups would still serve to test those modified CSL models.

In the case of the dissipative CSL model, the CSL-induced noise is linked to a ficticious finite-temperature environment that would also cause friction with respect to a distinguished reference frame \cite{smirne2015dissipative}. Specifically, one can associate to the CSL-induced momentum diffusion rate $D=(m/{1\,\rm u})^2\hbar^2\lambda/4  r_C^2$ per single atom of mass $m$ and per spatial direction a friction rate $\gamma=D/k_{\rm B}Tm$ based on the fluctuation-dissipation theorem. Reasonable CSL temperatures  of the order of the cosmic radiation background, say $T \simeq 1\,$K, are much higher in energy than the excitation quanta considered here, and so the finite temperature would add to the conventional CSL effect an effective damping on time scales $\gamma^{-1}$ without affecting the dephasing. However, the envisaged CSL rate parameters $\lambda \lesssim 10^{-10}\,$Hz and $r_C \simeq 10^{-7}\,$m yield $\gamma^{-1} \gtrsim 10^{13}\,$s, i.e.~no relevant effect on experimental time scales. Apart from that, the CSL-induced diffusion effect in the SWI scheme could be suppressed for $T \simeq \hbar\omega/k_B= 5\times10^{-7}\,$K. However, such low CSL temperatures have already been falsified in optomechanical heating experiments \cite{pontin2020ultranarrow,vinante2020testing}. 

The colored CSL model introduces a high-frequency cutoff in the CSL noise spectrum \cite{adler2007collapse,adler2008collapse}, which was pointed out to prohibit extreme events on the keV (or $10^{18}\,$Hz) range such as the spontaneous excitation of inner-shell electrons in atoms, leading to strong CSL bounds with x-ray detectors \cite{Arnquist2022Search}. In our case, a colored noise spectrum could impact the CSL-induced condensate depletion and diffusion effect, but only if the cutoff (implausibly) were as low as $10^5\,$Hz. On the other hand, the CSL-induced dephasing effect that affects both the DWI and the SWI scheme would remain unaltered.

\section{Conclusion}\label{Conclusion}
We presented experimental schemes based on two-mode atom BEC interference with either spatially separate or overlapping modes that are capable of testing spontaneous collapse proposed as a solution to the measurement problem. The suggested schemes rely solely on standard techniques such as squeezing and the manipulation of the interaction strength via e.g.~Feshbach resonances. Importantly, they neither require the preparation of maximally entangled GHZ states nor the detection of genuine $N$-atom correlations.

The unprecedented sensitivity scaling with the third power of the atom number in the single-well interferometer is an effect of the interplay between CSL-induced atom diffusion and interaction-induced phase dispersion. 
This should facilitate macroscopic quantum tests with precision atom interferometry, in laboratory or space-based experiments \cite{abou2020aedge}. 

In comparison to classical heating experiments, which currently provide the best CSL bounds, the here proposed interferometric test schemes are robust against conceivable modifications of collapse models with colored noise or friction. The latter confine the heating effect to a finite frequency window and temperature, while leaving the decoherence effect on macroscopic superpositions largely intact. 

Future work could explore interference protocols based on Bloch oscillations of atoms condensed in an optical lattice, instead of the single- or double-well traps considered here. Seconds-long coherence times with $N>10^4$ atoms were achieved in experiments \cite{Gustavsson2008,fattori2008magnetic}, which also demonstrated the Feshbach control over the atom-atom scattering length that facilitates switching between positive and negative phase dispersion.

\acknowledgements
We thank Naceur Gaaloul, Carsten Klempt, and Ernst M.\,Rasel for helpful discussions about BEC interferometry. B.\,S.\,is supported by Deutsche Forschungsgemeinschaft (DFG, German Research Foundation), Grant No. 449674892. P.\,H.\,thanks the Austrian Science Fund (FWF) Grant No. Y-1121, P 35953-N.


%

\end{document}